\documentclass[a4paper]{article}
\usepackage{amsmath,amssymb,bm}
\newtheorem{theorem}{Theorem}[section]
\newtheorem{lemma}{Lemma}[section]
\begin{document}
\begin{center}
\Large{ Ultradiscrete Pl\"ucker Relation Specialized for Soliton Solutions}\\
\vspace{1cm}
\large{Hidetomo Nagai\footnote{e-mail hdnagai@aoni.waseda.jp}, Daisuke Takahashi\footnote{e-mail daisuket@waseda.jp}}\\
Faculty of Science and Engineering, Waseda University, 3-4-1, Okubo, Shinjuku-ku, Tokyo 169-8555, Japan
\end{center}
\begin{abstract}
We propose an ultradiscrete analogue of Pl\"ucker relation specialized for soliton solutions.  It is expressed by an ultradiscrete permanent which is obtained by ultradiscretizing the permanent, that is, the signature-free determinant. Using this relation, we also show soliton solutions to the ultradiscrete KP equation and the ultradiscrete two-dimensional Toda lattice equation respectively.
\end{abstract}
\section{Introduction}
Soliton equations have been researched for several decades.  There are many equations expressed by different levels of discreteness.  Now we have continuous, semi-discrete, discrete and ultradiscrete soliton equations.  The continuous soliton equation is expressed by a partial differential equation and the semi-discrete soliton equation by a system of ordinary or partial differential equations.  The Kadomtsev-Petviashvili (KP) equation and the two-dimensional Toda lattice equation are continuous and semi-discrete respectively and they are fundamental for the soliton theory\cite{Freeman, 2DToda}.  These equations are transformed into bilinear forms, and their solutions are expressed by Wronski determinants.\par
  In general, soliton solutions in the determinant form obey Pl\"ucker relations and the relations are transformed into the soliton equations replacing the operations on the determinants by the differential or difference operators\cite{Sato, Ohta}.  This structure enables us to view the hierarchy and the common structure of soliton equations.  In fact, many soliton equations including the Korteweg-de Vries (KdV) equation, the Toda lattice equation and the sine-Gordon equation are obtained from the KP equation or the two-dimensional Toda lattice equation by the reduction of variables.\par
Discrete soliton equation is an equation of which independent variables are all discrete.  The discrete soliton equation is also expressed by the bilinear form and its determinant solution satisfies the Pl\"ucker relation.  In this case, the solution is expressed by the Casorati determinant.\par
  Ultradiscrete soliton equation is an equation of which all dependent and independent variables can take integer values.  It is derived from a discrete soliton equation by the ultradiscretization\cite{Tokihiro}, which is a limiting procedure of dependent variable using a key formula,
\begin{equation}
  \lim_{\varepsilon \to +0}\varepsilon \log (e^{a/\varepsilon}+e^{b/\varepsilon}) = \max (a, b).
\end{equation}
Ultradiscrete soliton equation has also soliton solutions\cite{Tsujimoto, Matsukidaira}.  Some interesting properties on the equation are discovered recently.  For instance, Nakamura discovered a soliton solution with a periodic phase for the ultradiscrete hungry Lotka-Volterra equation\cite{Nakamura}. Nakata proposed the vertex operator for the ultradiscrete KdV (uKdV) equation or the non-autonomous ultradiscrete KP (uKP) equation and showed their solutions\cite{Nakata, NakatauKP}.\par
  Moreover, the authors and Hirota proposed the ultradiscrete analogue of determinant solutions though the determinant cannot be ultradiscretized directly\cite{uKdV, Nagai, NagaiBack}.  Instead of the determinant, they used an ultradiscrete permanent (UP) defined by 
\begin{equation} 
  \max[a_{ij}]_{1\le i, j\le N}\equiv \max_{\pi} \sum_{1\le i\le N}a_{i\pi_i},
\end{equation}
where $a_{ij}$ is an arbitrary $N\times N$ matrix and $\pi=\{\pi_1,\pi_2,\ldots,\pi_N\}$ is an arbitrary permutation of 1, 2, $\dots$, $N$.  The soliton solutions in the UP form for the uKdV equation and the ultradiscrete Toda equation are shown in \cite{uKdV, Nagai}.  There exist B\"acklund transformations of ultradiscrete soliton equations\cite{NagaiBack}.\par
  The $(i, j)$ element of these UP soliton solutions is generally expressed by $|y_i+jr_i|$, where $y_i$ and $r_i$ are arbitrary parameters, and $|x|$ denotes an absolute value of $x$.  For example, the soliton solution to the uKdV equation is given by
\begin{equation}  
  f^n_i =\max 
\begin{bmatrix}
  |s_1(n, i)+2p_1| & |s_1(n, i)+4p_1| & \dots &|s_1(n, i)+2Np_1|\\
  \dots & \dots &  \dots &\dots\\
  |s_N(n, i)+2p_N| & |s_N(n, i)+4p_N| & \dots &|s_N(n, i)+2Np_N|
\end{bmatrix},
\end{equation}
where 
\begin{equation}
  s_j(n, i)=p_jn-q_ji+c_j \qquad q_j=\frac{1}{2} (|p_j+1|-|p_j-1|).  
\end{equation}
Though the expression of an ultradiscrete solution is analogous to that of discrete solution, we have not established the ultradiscretized Pl\"ucker relation.  Therefore, we have used the individual method to find the solution for every ultradiscrete soliton equation.\par
  This is due to the differences of basic operations between the determinant and the UP.  We show an example of such differences as follows. The determinant satisfy 
\begin{equation}  \label{determinant formula}
  \begin{vmatrix}
  a_{11} & a_{11}+a_{12}\\
  a_{21} & a_{21}+a_{22}  
  \end{vmatrix}
  =
  \begin{vmatrix}
  a_{11} & a_{12}\\
  a_{21} & a_{22}  
  \end{vmatrix}
  = a_{11}a_{22}-a_{12}a_{21}
\end{equation}
for any $a_{ij}$ $(1\le i, j\le 2)$.  When we consider the UP corresponding to the left-hand side of (\ref{determinant formula}), we have
\begin{equation}   \label{1-1}
  \max \begin{bmatrix}  
  a_{11} & \max(a_{11},a_{12}) \\
  a_{21} & \max(a_{21},a_{22})
  \end{bmatrix}.
\end{equation}
Then, using a property of UP
\begin{equation}  \label{property1}
\begin{aligned}
  &\max [\bm{b}_1 \ \dots \ \bm{b}_{j-1} \ \max(\bm{b}_j, \bm{b}'_j) \ \bm{b}_{j+1} \ \dots \ \bm{b}_N]\\
=&\max\bigl( 
  \max [\bm{b}_1 \ \dots \ \bm{b}_{j-1} \ \bm{b}_j \ \bm{b}_{j+1} \ \dots \ \bm{b}_N], \
  \max [\bm{b}_1 \ \dots \ \bm{b}_{j-1} \ \bm{b}'_j \ \bm{b}_{j+1} \ \dots \ \bm{b}_N]
\bigr),
\end{aligned}
\end{equation}
where $\bm {b}_j$ and $\bm {b'}_j$ $(1\le j\le N)$ are arbitrary $N$-dimensional vectors and 
$\max (\bm{b_j}, \bm {b'_j})$ denotes
\begin{equation}  \label{expression 1}
  \max (\bm{b_j}, \bm {b'_j}) \equiv \begin{pmatrix} \max (b_1, b'_1) \\ \max (b_2, b'_2)  \\ \dots \\ \max (b_N, b'_N)  \end{pmatrix},
\end{equation} we can expand (\ref{1-1}),
\begin{equation}
\begin{aligned}
  \max \begin{bmatrix}
  a_{11} & \max(a_{11},a_{12}) \\
  a_{21} & \max(a_{21},a_{22})
  \end{bmatrix}
  &= \max \left( 
 \max \begin{bmatrix}
  a_{11} & a_{11}\\
  a_{21} & a_{21} 
  \end{bmatrix}, \ 
 \max \begin{bmatrix}
  a_{11} & a_{12}\\
  a_{21} & a_{22} 
 \end{bmatrix} 
 \right) \\
  &=\max(a_{11}+a_{21}, a_{11}+a_{22}, a_{12}+a_{21}).  
\end{aligned}
\end{equation}
In contrast to the determinant case, the first argument in the right-hand side cannot be neglected.  Hence (\ref{1-1}) is not always equal to
\begin{equation}
  \max \begin{bmatrix}
  a_{11} & a_{12}\\
  a_{21} & a_{22}  
  \end{bmatrix},
\end{equation}
and it means UP does not have the relation such as (\ref{determinant formula}).\par
  The above kind of differences cause many troubles when we verify the solutions.  For example, one of the simplest Pl\"ucker relations is
\begin{equation}  \label{Plucker}
\begin{aligned}
  &|\bm{a}_1 \ldots \bm{a}_{N-1} \ \bm{b}_1| \times |\bm{a}_1 \ldots \bm{a}_{N-2} \ \bm{b}_2 \ \bm{b}_3|\\
  -&|\bm{a}_1 \ldots \bm{a}_{N-1} \ \bm{b}_2| \times |\bm{a}_1 \ldots \bm{a}_{N-2} \ \bm{b}_1 \ \bm{b}_3|\\
  +&|\bm{a}_1 \ldots \bm{a}_{N-1} \ \bm{b}_3| \times |\bm{a}_1 \ldots \bm{a}_{N-2} \ \bm{b}_1 \ \bm{b}_2|=0,
\end{aligned}
\end{equation}
for any $N$-dimensional column vectors $\bm{a}_j$ and $\bm{b}_j$.  However, the similar identity does not exist for the UP case.  Instead, Hirota showed UP's satisfy the following identity\footnote{Hirota gives an identity of ultradiscrete analogue of Pfaffian in \cite{Hirota-uhafnian}, and it reduces to (\ref{id of UP}) with proper conditions.  We give another proof in terms of UP in Appendix A. }\cite{Hirota-uhafnian}: 
\begin{equation}  \label{id of UP}
\begin{aligned}
 \max\bigl( &\max [\bm{a}_1 \ldots \bm{a}_{N-1} \ \bm{b}_1] + \max[ \bm{a}_1 \ldots \bm{a}_{N-2} \ \bm{b}_2 \ \bm{b}_3], \\
  &\max[\bm{a}_1  \ldots \bm{a}_{N-1} \ \bm{b}_2] + \max[\bm{a}_1 \ldots \bm{a}_{N-2} \ \bm{b}_1 \ \bm{b}_3]\bigr)\\
 =\max\bigl( &\max [\bm{a}_1 \ldots \bm{a}_{N-1} \ \bm{b}_1] + \max[ \bm{a}_1 \ldots \bm{a}_{N-2} \ \bm{b}_2 \ \bm{b}_3], \\
  &\max[\bm{a}_1  \ldots \bm{a}_{N-1} \ \bm{b}_3] + \max[\bm{a}_1 \ldots \bm{a}_{N-2} \ \bm{b}_1 \ \bm{b}_2]\bigr)\\
 =\max\bigl( &\max [\bm{a}_1 \ldots \bm{a}_{N-1} \ \bm{b}_2] + \max[ \bm{a}_1 \ldots \bm{a}_{N-2} \ \bm{b}_1 \ \bm{b}_3], \\
  &\max[\bm{a}_1 \ldots \bm{a}_{N-1} \ \bm{b}_3] + \max[\bm{a}_1 \ldots \bm{a}_{N-2} \ \bm{b}_1 \ \bm{b}_2]\bigr).
\end{aligned}
\end{equation}
This identity is not useful for the verification on ultradiscrete solutions since the anti-symmetry does not hold as shown in (\ref{determinant formula}) for determinants.\par
  In this article, we consider a general UP expression specialized for ultradiscrete soliton solutions.  The $(i,j)$ element of the specialized UP is defined by $|y_i+jr_i|$ where $y_i$ and $r_i$ are arbitrary constants.  Imposing this condition, we give a relation which corresponds to (\ref{Plucker}) in Section 2.  We call this relation the conditional ultradiscrete Pl\"ucker relation.  In Section 3 and 4, we present UP soliton solutions to the uKP equation and the ultradiscrete two-dimensional (u2D) Toda lattice equation respectively, and show that these solutions are verified by means of the conditional uPl\"ucker relation.  Finally, we give the concluding remarks in Section 5.\par
\section{Conditional ultradiscrete Pl\"ucker relation}
We give the following theorem in this section.  
\begin{theorem}  \label{theorem 1}
Let $\bm{x}_j $ be an $N$-dimensional vector defined by
\begin{equation}
  \bm{x}_j = 
\begin{pmatrix}
|y_1 +j r_1|\\ |y_2 +j r_2|\\ \dots \\ |y_N +j r_N|
\end{pmatrix}
\qquad \text{($y_i$, $r_i$ : arbitrary constants)}.  
\end{equation}
Then
\begin{equation}  \label{cond uPlucker N}
\begin{aligned}
  &\max[ \bm{x}_1 \ \dots \ \widehat{\bm{x}_{k_2}} \ \dots \ \bm{x}_{N+1} ] + \max[ \bm{x}_1 \ \dots \ \widehat{\bm{x}_{k_1}} \ \dots \ \widehat{\bm{x}_{k_3}} \ \dots \ \bm{x}_{N+2} ] \\
  =\max \bigl( & \max[ \bm{x}_1 \ \dots \ \widehat{\bm{x}_{k_3}} \ \dots \ \bm{x}_{N+1} ] + \max[ \bm{x}_1 \ \dots \ \widehat{\bm{x}_{k_1}}\ \dots \ \widehat{\bm{x}_{k_2}} \ \dots \ \bm{x}_{N+2} ], \\
  & \max[ \bm{x}_1 \ \dots \ \widehat{\bm{x}_{k_1}} \ \dots \ \bm{x}_{N+1} ] + \max[ \bm{x}_1 \ \dots \ \widehat{\bm{x}_{k_2}} \ \dots \ \widehat{\bm{x}_{k_3}}\ \dots \ \bm{x}_{N+2} ] \bigr)
\end{aligned}
\end{equation}
holds.  Here $1\le k_1<k_2<k_3\le N+1$ and the symbol $\widehat{\bm{x}_{k_j}}$ means that $\bm{x}_{k_j}$ is omitted.
\end{theorem}
Let us call (\ref{cond uPlucker N}) `conditional ultradiscrete Pl\"ucker(uPl\"ucker) relation'.  We note (\ref{cond uPlucker N}) can be rewritten as 
\begin{equation}  
\begin{aligned}
  &\max[M\ \bm{x}_{k_1} \ \bm{x}_{k_3} ] + \max[M\ \bm{x}_{k_2} \ \bm{x}_{N+2} ] \\
  =&\max \bigl(  \max[M\ \bm{x}_{k_1} \ \bm{x}_{k_2} ] + \max[M\ \bm{x}_{k_3} \ \bm{x}_{N+2} ], \ \max[M\ \bm{x}_{k_2} \ \bm{x}_{k_3} ] + \max[M\ \bm{x}_{k_1} \ \bm{x}_{N+2} ] \bigr)  
\end{aligned}
\end{equation}
with an $N\times (N-2)$ matrix $M$ defined by 
\begin{equation}
  M \equiv [\bm{x}_1 \ \dots \ \widehat{\bm{x}_{k_1}} \ \dots \ \widehat{\bm{x}_{k_2}} \ \dots \ \widehat{\bm{x}_{k_3}} \ \dots \ \bm{x}_{N+1}].  
\end{equation}
In order to prove Theorem \ref{theorem 1}, we give several lemmas.  
\begin{lemma}  \label{lemma1}
If an inequality 
\begin{equation} \label{remark1}
\begin{aligned}
  &\max[ \bm{x}_1 \ \dots \ \widehat{\bm{x}_{k_2}} \ \dots \ \bm{x}_{N+1} ] + \max[ \bm{x}_1 \ \dots \ \widehat{\bm{x}_{k_1}} \ \dots \ \widehat{\bm{x}_{k_3}} \ \dots \ \bm{x}_{N+2} ] \\
  \ge & \max[ \bm{x}_1 \ \dots \ \widehat{\bm{x}_{k_3}} \ \dots \ \bm{x}_{N+1} ] + \max[ \bm{x}_1 \ \dots \ \widehat{\bm{x}_{k_1}} \ \dots \ \widehat{\bm{x}_{k_2}} \ \dots \ \bm{x}_{N+2} ]
\end{aligned}
\end{equation}
holds, then (\ref{cond uPlucker N}) holds.  
\end{lemma}
\begin{lemma}  \label{lemma2}
The relation (\ref{cond uPlucker N}) can be rewritten as 
\begin{equation}  \label{remark2}
\begin{aligned}
  &\max[ \bm{x}_2 \ \dots \ \widehat{\bm{x}_{k_2}} \ \dots \ \bm{x}_{N+2} ] + \max[ \bm{x}_1 \ \dots \ \widehat{\bm{x}_{k_1}} \ \dots \ \widehat{\bm{x}_{k_3}} \ \dots \ \bm{x}_{N+2} ] \\
 = \max\bigl( & \max[ \bm{x}_2 \ \dots \ \widehat{\bm{x}_{k_3}} \ \dots \ \bm{x}_{N+2} ] + \max[ \bm{x}_1 \ \dots \ \widehat{\bm{x}_{k_1}} \ \dots \ \widehat{\bm{x}_{k_2}} \ \dots \ \bm{x}_{N+2} ],\\
  & \max[ \bm{x}_2 \ \dots \ \widehat{\bm{x}_{k_1}} \ \dots \ \bm{x}_{N+2} ] + \max[ \bm{x}_1 \ \dots \ \widehat{\bm{x}_{k_2}} \ \dots \ \widehat{\bm{x}_{k_3}} \ \dots \ \bm{x}_{N+2} ] \bigr), 
\end{aligned}
\end{equation}
where $1<k_1<k_2<k_3\le N+2$.  
\end{lemma}
\begin{lemma}  \label{lemma3}
If 
\begin{equation}
  0\le |r_1|\le |r_2| \le \dots \le |r_{N-1}|\le r_N, 
\end{equation}
then the $N$th-order UP can be reduced to the $(N-1)$th-order UP as
\begin{equation}  \label{remark3}
\begin{aligned}
  \max[ \bm{x}_{j_1} \ \bm{x}_{j_2}\ \dots \ \bm{x}_{j_N} ]
  =\max\bigl(  y_N+j_Nr_N+ \max[ \bm{\tilde x}_{j_1} \ \bm{\tilde x}_{j_2}\ \dots \ \bm{\tilde x}_{j_{N-1}} ],  \\
  -y_N-j_1r_N+ \max[ \bm{\tilde x}_{j_2} \ \bm{\tilde x}_{j_3}\ \dots \ \bm{\tilde x}_{j_N} ] \bigr),
\end{aligned}
\end{equation}
where $j_1<j_2<\dots <j_N$ and $\bm{\tilde x_j}$ denotes an $(N-1)$-dimensional vector
\begin{equation}
 \bm {\tilde x}_j = \begin{pmatrix}  |y_1 +j r_1|\\ |y_2 +j r_2|\\ \dots \\ |y_{N-1} +j r_{N-1}|  \end{pmatrix}.  
\end{equation}
\end{lemma}
Lemma \ref{lemma1} is derived from (\ref{id of UP}).  Lemma \ref{lemma2} is obtained since each $\bm {x} _j$ of (\ref{cond uPlucker N}) can be rewritten as $\bm{x}_{-j+N+3}$ with suitable transformations.  About Lemma \ref{lemma3}, the UP is expressed by
\begin{equation}  \label{eq:lemma3}
\begin{aligned}
  \max[ \bm{x}_{j_1} \ \bm{x}_{j_2}\ \dots \ \bm{x}_{j_N} ] =&\max_{\rho_i=\pm 1, \pi_i} \sum_{1\le i\le N}\rho_i(y_i+\pi_i r_i)\\
  =&\max_{\rho_i=\pm 1}\bigl( \sum_{1\le i\le N}\rho_iy_i+\max_{\pi_i}\sum_{1\le i\le N}\rho_i\pi_i r_i\bigr),
\end{aligned}
\end{equation}  
where $(\pi_1, \pi_2, \dots, \pi_N)$ denotes an arbitrary permutation of 1, 2, $\dots$, $N$.  The maximum of (\ref{eq:lemma3}) is given by $\pi_N=j_N$ in the case of $\rho_N=1$, and $\pi_N=j_1$ in the case of $\rho_N=-1$\cite{uKdV}.  Thus we obtain Lemma \ref{lemma3}.  \par
For Lemma \ref{lemma1}, Theorem \ref{theorem 1} is proved if we show (\ref{remark1}).  Then let us prove (\ref{remark1}) with a mathematical induction.  Hereafter, we adopt a simple notation $j$ for $\bm{x}_j$.  For $N=2$, one can prove 
\begin{equation}  
  \max[1 \ 3] + \max[ 2 \ 4] \ge  \max [1 \ 2] + \max[ 3 \ 4].
\end{equation}
Then let us show the inequality 
\begin{equation}  \label{2-2}
\begin{aligned}
  &\max[ 1 \ \dots \ \widehat{k_2} \ \dots \ N+2 ] + \max[ 1 \ \dots \ \widehat{k_1} \ \dots \ \widehat{k_3} \ \dots \ N+3 ] \\
  \ge & \max[ 1 \ \dots \ \widehat{k_3} \ \dots \ N+2 ] + \max[ 1 \ \dots \ \widehat{k_1} \ \dots \ \widehat{k_2} \ \dots \ N+3 ]
\end{aligned}
\end{equation}
for $1\le k_1<k_2<k_3\le N+2$ under the assumptions (\ref{remark1}) and 
\begin{equation}  \label{2-2-2}
  0\le |r_1|\le |r_2|\le \dots \le |r_N|.
\end{equation}
We note (\ref{2-2-2}) can be assumed without loss of generality.  \par
In the case of $1<k_1<k_2<k_3<N+2$ and $r_{N+1}>|r_N|$, the UP's of the left-hand side in (\ref{2-2}) are rewritten as
\begin{equation} \label{2-3-1}
\begin{aligned}
  \max[ 1 \ \dots \ \widehat{k_2} \ \dots \ N+2 ] 
  =\max \bigl( &y_{N+1} +(N+2)r_{N+1}+\max[ 1 \ \dots \ \widehat{k_2} \ \dots \ N+1 ] , \\
  &-y_{N+1}-r_{N+1} +\max[ 2 \ \dots \ \widehat{k_2} \ \dots \ N+2 ] \bigr),
\end{aligned}
\end{equation}
and 
\begin{equation} \label{2-3-2}
\begin{aligned}
  \max[ 1 \ \dots \ \widehat{k_1} \ \dots \ \widehat{k_3} \dots \ N+3 ] =\max \bigl( &y_{N+1}+(N+3)r_{N+1}+\max[ 1 \ \dots \ \widehat{k_1} \ \dots \ \widehat{k_3} \dots \ N+2 ], \\
  &-y_{N+1}-r_{N+1}+\max[ 2 \ \dots \ \widehat{k_1} \ \dots \ \widehat{k_3} \dots \ N+3 ]\bigr),
\end{aligned}
\end{equation}
respectively by Lemma \ref{lemma3}.  Therefore, a sum of (\ref{2-3-1}) and (\ref{2-3-2}) is expressed by
\begin{equation}  \label{left}
\begin{aligned}
  &\max[ 1 \ \dots \ \widehat{k_2} \ \dots \ N+2 ]+ \max[ 1 \ \dots \ \widehat{k_1} \ \dots \ \widehat{k_3} \dots \ N+3 ] \\
  =\max \bigl( &2y_{N+1} +(2N+5)r_{N+1} +\max[ 1 \ \dots \ \widehat{k_2} \ \dots \ N+1 ]+ \max[ 1 \ \dots \ \widehat{k_1} \ \dots \ \widehat{k_3} \dots \ N+2 ], \\
  & -2y_{N+1} -2r_{N+1} + \max[ 2 \ \dots \ \widehat{k_2} \ \dots \ N+2 ] +\max[ 2 \ \dots \ \widehat{k_1} \ \dots \ \widehat{k_3} \dots \ N+3 ], \\
  & (N+2)r_{N+1} + \max[ 2 \ \dots \ \widehat{k_2} \ \dots \ N+2 ]+ \max[ 1 \ \dots \ \widehat{k_1} \ \dots \ \widehat{k_3} \dots \ N+2 ],\\
  &(N+1)r_{N+1} +\max[ 1 \ \dots \ \widehat{k_2} \ \dots \ N+1 ] + \max[ 2 \ \dots \ \widehat{k_1} \ \dots \ \widehat{k_3} \dots \ N+3 ]\bigr).
\end{aligned}
\end{equation} 
Similarly, the right-hand side in (\ref{2-2}) is expressed by
\begin{equation} \label{right}
\begin{aligned}
  \max \bigl( &2y_{N+1} +(2N+5)r_{N+1} +\max[ 1 \ \dots \ \widehat{k_3} \ \dots \ N+1 ]+ \max[ 1 \ \dots \ \widehat{k_1} \ \dots \ \widehat{k_2} \ \dots \ N+2 ], \\
  & -2y_{N+1} -2r_{N+1} + \max[ 2 \ \dots \ \widehat{k_3} \ \dots \ N+2 ] +\max[ 2 \ \dots \ \widehat{k_1} \ \dots \ \widehat{k_2} \ \dots \ N+3 ], \\
  & (N+2)r_{N+1} + \max[ 2 \ \dots \ \widehat{k_3} \ \dots \ N+2 ]+ \max[ 1 \ \dots \ \widehat{k_1} \ \dots \ \widehat{k_2} \ \dots \ N+2 ], \\
  & (N+1)r_{N+1} +\max[ 1 \ \dots \ \widehat{k_3} \ \dots \ N+1 ] + \max[ 2 \ \dots \ \widehat{k_1} \ \dots \ \widehat{k_2} \ \dots \ N+3 ] \bigr).  
\end{aligned}
\end{equation} 
The first and second arguments of (\ref{left}) in the right-hand side are greater than those of (\ref{right}) from the assumption.  The third argument of (\ref{left}) in the right-hand side is also greater than that of (\ref{right}) from Lemma \ref{lemma2}.  Moreover, the following lemma holds.  
\begin{lemma}  \label{lemma4}
Inequalities 
\begin{equation} \label{ineq1}
\begin{aligned}
   r_{N+1} + &\max[ 2 \ \dots \ \widehat{k_2} \ \dots \ N+2 ]+ \max[ 1 \ \dots \ \widehat{k_1} \ \dots \ \widehat{k_3} \dots \ N+2 ]\\
  \ge &\max[ 1 \ \dots \ \widehat{k_2} \ \dots \ N+1 ] + \max[ 2 \ \dots \ \widehat{k_1} \ \dots \ \widehat{k_3} \dots \ N+3 ]
\end{aligned}
\end{equation} 
and
\begin{equation} \label{ineq2}
\begin{aligned}
   r_{N+1} + &\max[ 2 \ \dots \ \widehat{k_3} \ \dots \ N+2 ]+ \max[ 1 \ \dots \ \widehat{k_1} \ \dots \ \widehat{k_2} \ \dots \ N+2 ] \\
  \ge & \max[ 1 \ \dots \ \widehat{k_3} \ \dots \ N+1 ] + \max[ 2 \ \dots \ \widehat{k_1} \ \dots \ \widehat{k_2} \ \dots \ N+3 ]    
\end{aligned}
\end{equation} 
hold for $1<k_1<k_2<k_3<N+2$.  
\end{lemma}
Lemma \ref{lemma4} is proved by a mathematical induction shown in Appendix B.  
Thus, the fourth argument is smaller than the third one in (\ref{left}) and (\ref{right}) respectively.  Therefore, (\ref{2-2}) holds in the case of $1<k_1<k_2<k_3<N+2$ and $r_{N+1}>|r_N|$.  The similar procedure enable us to prove in the other cases. Hence, we obtain the conditional uPl\"ucker relation.  
\section{The ultradiscrete KP equation and its UP solution}
Let us consider the following tau function defined by UP. 
\begin{equation} \label{sol uKP}
  \tau(l, m, n) = \max [ \phi_i (l, m, n, s+j-1)]_{1\le i, j\le N} ,
\end{equation}
where $s$ is an auxiliary variable, and $\phi_i(l, m, n, s)$ is defined by
\begin{equation} \label{def phi of uKP}
  \phi_i(l, m, n, s) =\max(\eta_i(l, m, n, s), \eta'_i(l, m, n, s))
\end{equation}
with
\begin{equation} \label{def eta of uKP}
\begin{aligned}
  &\eta_i(l, m, n, s)= p_i s+\max (0, p_i-a_1)l+\max (0, p_i-a_2)m+\max (0, p_i-a_3)n+c_i,\\
  &\eta'_i(l, m, n, s)=-p_i s+\max (0, -p_i-a_1)l+\max (0, -p_i-a_2)m+\max (0, -p_i-a_3)n+c'_i.
\end{aligned}
\end{equation}
Here $a_1$, $a_2$ and $a_3$ are parameters satisfying $a_1>a_2>a_3$, and $p_i$, $c_i$ and $c'_i$ are arbitrary parameters.  One can obtain the following relations:
\begin{align} 
  &\phi_i(l+1, m, n, s) = \max( \phi_i(l, m, n, s), \ \phi_i(l, m, n, s+1)-a_1),  \label{uKP cond1} \\
  &\phi_i(l, m+1, n, s) = \max( \phi_i(l, m, n, s), \ \phi_i(l, m, n, s+1)-a_2),  \label{uKP cond1-2} \\
  &\phi_i(l, m, n+1, s) = \max( \phi_i(l, m, n, s), \ \phi_i(l, m, n, s+1)-a_3)  \label{uKP cond1-3}
\end{align}
and 
\begin{equation} \label{uKP cond2}
\begin{aligned}
  \phi_{i_1}(l, m, n, s+j)+\phi_{i_2}(l, m, n, s+j)
\le \max ( &\phi_{i_1}(l, m, n, s+j-1) +\phi_{i_2}(l, m, n, s+j+1), \\
  &  \phi_{i_2}(l, m, n, s+j-1) +\phi_{i_1}(l, m, n, s+j+1)) 
\end{aligned}
\end{equation}
for $1\le i, i_1, i_2\le N$.  We first rewrite the tau function with (\ref{uKP cond1}), (\ref{uKP cond1-2}), (\ref{uKP cond1-3}) and (\ref{uKP cond2}) in Subsection 3.1.  Second we give the relation shown by the conditional uPl\"ucker relation in Subsection 3.2. Finally, in Subsection 3.3, we give the UP solution for the uKP equation.
\subsection{Rewriting the tau function}
Using (\ref{uKP cond1}),  $\tau (l+1, m, n)$ is expanded as
\begin{equation}  \label{3-1}
\begin{aligned}
  \tau(l+1, m, n)&=\max [\phi_i (l+1, m, n, s+j-1)]_{1\le i, j\le N}\\
  &=\max [\max( \phi_i(l, m, n, s+j-1), \ \phi_i(l, m, n, s+j)-a_1 )]_{1\le i, j\le N}
\end{aligned}
\end{equation}
In particular, using the simple notations, 
\begin{equation}
  \begin{pmatrix} \phi_1(l, m, n, s+j) \\ \phi_2(l, m, n, s+j) \\ \dots \\ \phi_N(l, m, n, s+j) \end{pmatrix}  \equiv \begin{pmatrix} \phi_1(j) \\ \phi_2(j) \\ \dots \\ \phi_N(j) \end{pmatrix} \equiv \bm{\phi }(j) ,
\end{equation}
(\ref{3-1}) is expressed by
\begin{equation}  \label{3-1-2}
\begin{aligned}
  \tau(l+1, m, n)  &=\max [\max( \bm{\phi}(j-1), \ \bm{\phi}(j)-a_1 \cdot \bm{1})]_{1\le j\le N},
\end{aligned}
\end{equation}
where $\bm{1}$ and $\max (\bm{\phi} (j-1), \bm{\phi} (j))$ denote 
\begin{equation}
  \bm{1} \equiv \begin{pmatrix} 1 \\ 1 \\ \dots \\ 1\end{pmatrix}
\end{equation}
and (\ref{expression 1}) respectively.  Furthermore, by applying a property of UP (\ref{property1}) to each column in (\ref{3-1-2}), $\tau(l+1, m, n)$ is expanded as the maximum of the following $2^N$ UP's,
\begin{equation}
\begin{aligned}
  &\max [\bm{\phi}(0) \quad \bm{\phi}(1) \quad \bm{\phi}(2) \quad  \dots \quad \bm{\phi}(N-1)], \\ 
  &\max [\bm{\phi}(1)-a_1 \cdot \bm{1} \quad \bm{\phi}(1) \quad \bm{\phi}(2) \quad \dots \quad \bm{\phi}(N-1)], \\
  &\max [\bm{\phi}(0) \quad \bm{\phi}(2)-a_1 \cdot \bm{1} \quad \bm{\phi}(2) \quad \dots \quad \bm{\phi}(N-1)], \\
  &  \dots \\
  & \max [\bm{\phi}(1) -a_1 \cdot \bm{1} \quad \bm{\phi}(2) -a_1 \cdot \bm{1} \quad \bm{\phi}(3) -a_1 \cdot \bm{1} \quad \dots \quad \bm{\phi}(N) -a_1 \cdot \bm{1}].
\end{aligned}
\end{equation}
Let us call a set of the above UP's $S$.  Moreover, using another property of UP, 
\begin{equation}  \label{property2}
  \max [\bm{b}_1 \ \dots \ \bm{b}_{j-1} \quad \bm{b}_j +c \cdot \bm{1} \quad \bm{b}_{j+1} \ \dots \ \bm{b}_N]= \max [\bm{b}_1 \ \dots \ \bm{b}_{j-1} \ \ \bm{b}_j \ \ \bm{b}_{j+1} \ \dots \ \bm{b}_N]+c,  
\end{equation}
where $\bm{b}_j$ $(1\le j\le N)$ is an arbitrary $N$-dimensional vector and $c$ arbitrary constant, we can divide $S$ into $N+1$ sets as 
\begin{equation}
  S= \{ S_0, S_1-a_1, S_2-2a_1, \dots , S_N-Na_1 \}.
\end{equation}
For example, $S_0$ is expressed by 
\begin{equation}
  S_0=\{ \max [0 \ 1 \ 2 \ \dots \ N-1]\}
\end{equation}
where $j$ denotes $\bm{\phi}(j)$, and $S_1$ is 
\begin{equation}  \label{3-2}
S_1= \{ \max [1 \ 1 \ 2 \ \dots \ N-1], \ \max [0 \ 2 \ 2 \ 3 \ \dots \ N-1], \ \dots , \  \max [0 \ 1 \ 2 \ \dots \ N-2 \ N] \}. 
\end{equation}
About these sets of UP's, we give the following lemma.  
\begin{lemma}  \label{lemma3-1}
  An inequality 
\begin{equation}  \label{3-3}
  \max [M \ j \ j] \le \max [M \ j-1 \ j+1]
\end{equation}
holds for any $j$, where $M$ denotes an arbitrary $N\times (N-2)$ matrix.  
\end{lemma}
Lemma \ref{lemma3-1} is proved since each UP is expanded as
\begin{equation}  
\begin{aligned}
  &\max [M \ j \ j] =\max_{{1\le i_1, i_2\le N}\atop {i_1\not =i_2}}\Bigl( \max [M \ j \ j]_{ i_1,  i_2\atop{ N-1, N}} +\phi_{i_1}(j)+\phi_{i_2}(j)\Bigr),\\
  &\max [M \ j-1 \ j+1]=\max_{{1\le i_1, i_2\le N}\atop {i_1\not =i_2}}\Bigl( \max [M \ j-1 \ j+1]_{i_1, i_2 \atop{N-1, N}} +\phi_{i_1}(j-1)+\phi_{i_2}(j+1)\Bigr), 
\end{aligned}
\end{equation}
where $\max A _{i_1, i_2 \atop{N-1, N}}$ denotes the $(N-2)$th-order UP obtained by eliminating the $i_1$-th and $i_2$-th rows and the $(N-1)$-th and $N$-th columns from $N\times N$ matrix $A$.  Inequality (\ref{3-3}) is derived from 
\begin{equation}  
  \max [M \ j \ j]_{i_1, i_2\atop{N-1, N}} =\max [M \ j-1 \ j+1]_{i_1, i_2\atop{ N-1, N}}
\end{equation}
and (\ref{uKP cond2}).  \par
Therefore, $\max S_1$ is determined as $\max [0 \ 1 \ 2 \ \dots \ N-2 \ N]$ since 
\begin{equation}
\begin{aligned}
  &\max [1 \ 1 \ 2 \ \dots \ N-1] \\
  \le &\max [0 \ 2 \ 2 \ 3 \ \dots \ N-1] \\
  \le & \dots \\
  \le &\max [0 \ 1 \ 2 \ \dots \ N-1 \ N-1]\\
  \le &\max [0 \ 1 \ 2 \ \dots \ N-2 \ N].
\end{aligned}
\end{equation}
holds.  Similarly, other $\max S_{k_1}$ $(0\le k_1\le N)$ are determined, and $\tau(l+1, m, n)$ is reduced to the maximum of $(N+1)$ UP's and we obtain the following lemma.  
\begin{lemma}  \label{lemma3-2}
Tau function $\tau (l+1, m, n)$ is reduced to 
\begin{equation}  \label{tau(l+1,m,n)}  
  \tau(l+1, m, n) =\max_{0\le k_1\le N}( \tau_c(N-k_1, N+1)-k_1 a_1 ),  
\end{equation}
where $\tau_c(\alpha , \beta )$ $(\alpha <\beta )$ is the UP defined by
\begin{equation} 
\tau_c(\alpha, \beta ) = \max [0 \ \dots \ \widehat{\alpha} \ \dots \ \widehat{\beta} \ \dots \ N+1].  
\end{equation}
\end{lemma}
Furthermore, using (\ref{uKP cond1-2}) and (\ref{uKP cond1-3}) respectively, $\tau(l, m+1, n+1)$ is also reduced to the maximum of $(N+1)^2$ UP's as follows.  
\begin{lemma}  \label{lemma3-3}
Tau function $\tau(l, m+1, n+1)$ is reduced to  
\begin{equation}
  \tau(l, m+1, n+1)  =\max_{0\le k_2, k_3\le N}( \Psi (k_2, k_3)-k_2a_2-k_3a_3 ),
\end{equation}
where $\Psi(k_2, k_3)$ is defined by 
\begin{equation}  \label{KP Psi1}
  \Psi(k_2, k_3) =
 \begin{cases}
  \displaystyle \max_{0\le i\le N-k_3}(\tau_c(N-k_3-i, N-k_2+1+i)) & (k_3\ge k_2, N-k_2)\\
  \displaystyle  \max_{0\le i\le k_2}(\tau_c(N-k_2-k_3+i, N+1-i))
 & (N-k_2\ge k_3\ge k_2 )\\
  \displaystyle \max_{0\le i\le N-k_2}(\tau_c(N-k_2-i, N-k_3+1+i)) & (k_2\ge k_3\ge N-k_2)\\
  \displaystyle \max_{0\le i\le k_3}(\tau_c(N-k_2-k_3+i, N+1-i)) & (k_2, N-k_2\ge k_3)
\end{cases}.
\end{equation}
for $0\le k_2, k_3\le N$.  Especially, (\ref{KP Psi1}) gives   
\begin{equation}  \label{KP Psi2}
\begin{aligned}
  &\Psi(k_2-1, k_3) = \max (\Psi(k_2, k_3-1), \ \tau_c(N-k_3+1, N-k_2+1)), \qquad &(k_2>k_3 ) \\
  &\Psi(k_2-1, k_3) = \Psi(k_2, k_3-1), \qquad &(k_2=k_3 ) \\
  &\max( \Psi(k_2-1, k_3), \tau_c(N-k_2+1, N-k_3+1) )= \Psi(k_2, k_3-1), \qquad &(k_2<k_3 )
\end{aligned}
\end{equation}
for $1\le k_2, k_3\le N$.    
\end{lemma}
The proof of Lemma \ref{lemma3-3} is shown in Appendix C.  We can obtain the similar expressions for $\tau (l, m+1, n)$, $\tau(l, m, n+1)$, $\tau(l+1, m, n+1)$ and $\tau(l+1, m+1, n)$.  
\subsection{Identity for $\tau_c$}
About the function $\tau _c$, the following identity holds.  
\begin{equation}  \label{uPlucker for uKP}
  \tau_c(k_2, N+1) +  \tau_c (k_1, k_3) =\max ( \tau_c(k_1, N+1) +  \tau_c (k_2, k_3), \  \tau_c(k_3, N+1) + \tau_c (k_1, k_2) ),
\end{equation}
where $0< k_1 < k_2< k_3 <N+1$.  It is proved as below.  Equation (\ref{uPlucker for uKP}) is rewritten by 
\begin{equation}  \label{uKP uPlucker}
\begin{aligned}
  &\max[ \bm{\phi}(0) \ \dots \ \widehat{\bm{\phi}(k_2)} \ \dots \ \bm{\phi}(N) ] + \max[ \bm{\phi}(0) \ \dots \ \widehat{\bm{\phi}(k_1)} \ \dots \ \widehat{\bm{\phi}(k_3)} \ \dots \ \bm{\phi}(N+1) ] \\
  =\max \bigl( & \max[ \bm{\phi}(0) \ \dots \ \widehat{\bm{\phi}(k_3)} \ \dots \ \bm{\phi}(N) ] + \max[ \bm{\phi}(0) \ \dots \ \widehat{\bm{\phi}(k_1)}\ \dots \ \widehat{\bm{\phi}(k_2)} \ \dots \ \bm{\phi}(N+1) ], \\
  & \max[ \bm{\phi}(0) \ \dots \ \widehat{\bm{\phi}(k_1)} \ \dots \ \bm{\phi}(N) ] + \max[ \bm{\phi}(0) \ \dots \ \widehat{\bm{\phi}(k_2)} \ \dots \ \widehat{\bm{\phi}(k_3)}\ \dots \ \bm{\phi}(N+1) ] \bigr).
\end{aligned}
\end{equation}
Especially, let us recall the definition of $\bm{\phi}(j)$,
\begin{equation}
  \bm{\phi}(j) = \begin{pmatrix}
\max(\eta_1 +j p_1, \eta'_1-jp_1) \\ \max(\eta_2 +j p_2, \eta'_2-jp_2) \\ \dots \\ \max(\eta_N +j p_N, \eta'_N-jp_N)
\end{pmatrix},
\end{equation}
where $\eta_i$ and $\eta '_i$ denote $\eta_i(l, m, n, s)$ and $\eta '_i(l, m, n, s)$ for short.  By adding $\sum_{1\le i\le N}(-\eta_i-\eta'_i)/2$ to both sides in (\ref{uKP uPlucker}), it is reduced to the conditional uPl\"ucker relation, hence, proved.  
\subsection{Equations for the tau functions}
Substituting the expression of tau functions into  
\begin{equation} \label{uKP2 left}
\begin{aligned}
  \max (&\tau(l+1, m, n)+\tau(l, m+1, n+1)-a_1-a_2, \\
  &\tau(l, m+1, n)+\tau(l+1, m, n+1)-a_2-a_3, \\
  &\tau(l, m, n+1)+\tau(l+1, m+1, n)-a_1-a_3 )
\end{aligned}
\end{equation}
and
\begin{equation} \label{uKP2 right}
\begin{aligned}
  \max( &\tau(l+1, m, n)+\tau(l, m+1, n+1)-a_1-a_3, \\
   &\tau(l, m+1, n)+\tau(l+1, m, n+1)-a_1-a_2, \\
  &\tau(l, m, n+1)+\tau(l+1, m+1, n)-a_2-a_3 )
\end{aligned}
\end{equation}
respectively, we obtain 
\begin{equation} \label{uKP3 left}
\begin{aligned}
  \max_{0\le k_1, k_2, k_3\le N}\bigl( & \tau_c(N-k_1, N+1)  + \Psi (k_2, k_3) -(k_1+1)a_1-(k_2+1) a_2 -k_3 a_3 , \\
  &\tau_c(N-k_2, N+1) + \Psi (k_1, k_3) -k_1 a_1-(k_2+1)a_2  -(k_3+1) a_3, \\
  & \tau_c(N-k_3, N+1) + \Psi (k_1, k_2) -(k_1+1) a_1 -k_2 a_2-(k_3+1)a_3  \bigr) \\  
\end{aligned}
\end{equation}
and
\begin{equation} \label{uKP3 right}
\begin{aligned}
  \max_{0\le k_1, k_2, k_3\le N}\bigl( & \tau_c(N-k_1, N+1) + \Psi (k_2, k_3) -(k_1+1)a_1 -k_2 a_2 -(k_3+1) a_3, \\
  & \tau_c(N-k_2, N+1) + \Psi (k_1, k_3) -(k_1+1) a_1 -(k_2+1)a_2 -k_3 a_3, \\
  & \tau_c(N-k_3, N+1) + \Psi (k_1, k_2) -k_1 a_1 -(k_2+1) a_2-(k_3+1)a_3 \bigr).
\end{aligned}
\end{equation}
Let us show that (\ref{uKP3 left}) is equal to (\ref{uKP3 right}).  For this purpose, we compare the arguments which have the same $-k_1a_1-k_2a_2-k_3a_3$ in both.  \par
In the case of $k_1=0$, the argument in (\ref{uKP3 left}) is expressed by
\begin{equation}
  \tau_c(N-k_2, N+1)+\Psi (0, k_3)-(k_2+1)a_2  -(k_3+1) a_3. 
\end{equation}
On the other hand, that in (\ref{uKP3 right}) is expressed by
\begin{equation}
  \tau_c(N-k_3, N+1)+\Psi (0, k_2) -(k_2+1) a_2-(k_3+1)a_3. 
\end{equation}
They are equivalent for (\ref{KP Psi1}).  Similarly, if $k_2=0$ or $k_3=0$, then the arguments are equivalent.  \par
Next, we consider in the case of $k_1=N+1$.  When $k_2$ or $k_3$ is also $N+1$, both are obviously equivalent.  When $1\le k_2, k_3\le N$, each argument is expressed by
\begin{equation}  \label{uKP l-1}
\begin{aligned}
  \max\bigl(& \tau_c(0, N+1) + \Psi (k_2-1, k_3) -(N+1)a_1-k_2 a_2 -k_3 a_3 , \\
  & \tau_c(N-k_3+1, N+1) + \Psi (N, k_2)  -(N+1)a_1-k_2 a_2-k_3a_3\bigr) ,  
\end{aligned}
\end{equation}
\begin{equation}  \label{uKP r-1}
\begin{aligned}
  \max  \bigl( &\tau_c(0, N+1) + \Psi (k_2, k_3-1) -(N+1)a_1-k_2 a_2 -k_3 a_3, \\
  & \tau_c(N-k_2+1, N+1) + \Psi (N, k_3)  -(N+1)a_1-k_2a_2-k_3 a_3\bigr)
\end{aligned}
\end{equation}
respectively.  It is trivial that they coincide when $k_2=k_3$.  When $k_2>k_3$, (\ref{uKP l-1}) and (\ref{uKP r-1}) reduce to
\begin{equation}  \label{uKP l-2}
\begin{aligned}
  \max \bigl(& \tau_c(0, N+1) + \max( \Psi (k_2, k_3-1), \tau_c(N-k_3+1, N-k_2+1)), \\
  & \tau_c(N-k_3+1, N+1) + \tau_c (0, N-k_2+1)\bigr) -(N+1)a_1-k_2 a_2 -k_3 a_3, 
\end{aligned}
\end{equation}
\begin{equation}  \label{uKP r-2}
\begin{aligned}
  \max \bigl( & \tau_c(0, N+1) + \Psi (k_2, k_3-1), \\
  &\tau_c(N-k_2+1, N+1) + \tau_c (0, N-k_3+1) \bigr)-(N+1)a_1-k_2 a_2 -k_3 a_3
\end{aligned}
\end{equation}
for (\ref{KP Psi1}) and (\ref{KP Psi2}).  They also coincide since  
\begin{equation}  \label{uKP N+1}
\begin{aligned}
  &\max \bigl( \tau_c(0, N+1) + \tau_c(N-k_3+1, N-k_2+1), \tau_c(N-k_3+1, N+1) + \tau_c (0, N-k_2+1) \bigr) \\  
  =& \tau_c(N-k_2+1, N+1) + \tau_c (0, N-k_3+1),   
\end{aligned}
\end{equation}
holds for $1\le k_3< k_2\le N$ because of (\ref{uPlucker for uKP}).   It is also shown in the case of $k_2<k_3$.  \par
Finally, we consider in the case of $1\le k_1, k_2,  k_3 \le N$.  The arguments in (\ref{uKP3 left}) and (\ref{uKP3 right}) are expressed by
\begin{equation} 
\begin{aligned}
  \max (& \tau_c(N-k_1+1, N+1)+ \Psi (k_2-1, k_3), \\
  &\tau_c(N-k_2+1, N+1) + \Psi (k_1, k_3-1) , \\
  &\tau_c(N-k_3+1, N+1) + \Psi (k_1-1, k_2)),   
\end{aligned}
\end{equation}
\begin{equation} 
\begin{aligned}
  \max (& \tau_c(N-k_1+1, N+1) + \Psi (k_2, k_3-1), \\
  &\tau_c(N-k_2+1, N+1) + \Psi (k_1-1, k_3), \\
  &\tau_c(N-k_3+1, N+1) + \Psi (k_1, k_2-1)).
\end{aligned}
\end{equation}
It is clear that both correspond if $k_i=k_j$ ($i, j=1, 2, 3$ and $i\not=j$).  Then, we assume $k_1>k_2>k_3$ and have 
\begin{equation} 
\begin{aligned}
  \max (& \tau_c(N-k_1+1, N+1) + \max(\Psi(k_2, k_3-1), \tau_c (N-k_2+1, N-k_3+1)), \\
  & \tau_c(N-k_2+1, N+1) + \Psi (k_1, k_3-1), \\
  & \tau_c(N-k_3+1, N+1) + \max(\Psi(k_1, k_2-1), \tau_c (N-k_1+1, N-k_2+1)) ),
\end{aligned}
\end{equation}
\begin{equation} 
\begin{aligned}
  \max (& \tau_c(N-k_1+1, N+1)+ \Psi (k_2, k_3-1), \\ 
  &\tau_c(N-k_2+1, N+1) + \max(\Psi(k_1, k_3-1), \tau_c (N-k_1+1, N-k_3+1)), \\
  &\tau_c(N-k_3+1, N+1) + \Psi (k_1, k_2-1)).  
\end{aligned}
\end{equation}
They coincide since 
\begin{equation} 
\begin{aligned}
  \max ( &\tau_c(N-k_1+1, N+1) +  \tau_c (N-k_2+1, N-k_3+1),\\
  & \tau_c(N-k_3+1, N+1) + \tau_c (N-k_1+1, N-k_2+1) )\\
  =&\tau_c(N-k_2+1, N+1) +  \tau_c (N-k_1+1, N-k_3+1).
\end{aligned}
\end{equation}
holds by (\ref{uPlucker for uKP}).  \par
Therefore, we obtain the following lemma.  
\begin{lemma}
    The UP (\ref{sol uKP}) defined by (\ref{def phi of uKP}) and (\ref{def eta of uKP}) satisfies the equation,  
\begin{equation} \label{uKP2}
\begin{aligned}
  \max (&\tau(l+1, m, n)+\tau(l, m+1, n+1)-a_1-a_2, \\
   &\tau(l, m+1, n)+\tau(l+1, m, n+1)-a_2-a_3, \\
  &\tau(l, m, n+1)+\tau(l+1, m+1, n)-a_1-a_3 )\\
  =\max( &\tau(l+1, m, n)+\tau(l, m+1, n+1)-a_1-a_3, \\
   &\tau(l, m+1, n)+\tau(l+1, m, n+1)-a_1-a_2, \\
  &\tau(l, m, n+1)+\tau(l+1, m+1, n)-a_2-a_3 ).  
\end{aligned}
\end{equation}
\end{lemma}
In particular, it can be reduced to the uKP equation\cite{Shinzawa, NakatauKP},
\begin{equation} \label{uKP}
\begin{aligned}
  &\tau(l, m+1, n)+\tau(l+1, m, n+1)-a_2  \\
  =& \max (\tau(l+1, m, n)+\tau(l, m+1, n+1)-a_1, \tau(l, m, n+1)+\tau(l+1, m+1, n)-a_2)
\end{aligned}
\end{equation}   
since
\begin{equation}
\begin{aligned}
  &\tau(l+1, m, n)+\tau(l, m+1, n+1)-a_1-a_2<  \tau(l+1, m, n)+\tau(l, m+1, n+1)-a_1-a_3, \\
  &\tau(l, m+1, n)+\tau(l+1, m, n+1)-a_2-a_3>  \tau(l, m+1, n)+\tau(l+1, m, n+1)-a_1-a_2, \\
  &\tau(l, m, n+1)+\tau(l+1, m+1, n)-a_1-a_3<  \tau(l, m, n+1)+\tau(l+1, m+1, n)-a_2-a_3 
\end{aligned}
\end{equation}
hold for $a_1>a_2>a_3$.  We obtain therefore Theorem \ref{theorem 3}.  
\begin{theorem}  \label{theorem 3}
  The UP (\ref{sol uKP}) defined by (\ref{def phi of uKP}) and (\ref{def eta of uKP}) satisfies the uKP equation (\ref{uKP}).  
\end{theorem}  
\section{The ultradiscrete 2D Toda lattice equation and its UP solution}
In this section, we give the UP soliton solution to the u2D Toda lattice equation\cite{2DToda}, 
\begin{equation} \label{u2DToda}
\begin{aligned}
  \tau(l, m-1, n)+\tau(l+1, m, n)= &\max( \tau(l, m, n)+\tau(l+1, m-1, n), \\
  &\qquad \tau(l, m-1, n+1)+\tau(l+1, m, n-1)-\delta-\varepsilon ),  
\end{aligned}
\end{equation}
where $\delta, \varepsilon > 0$.  The procedure is similar to the previous section.  We only show the points of the proof.  \par 
Considering the tau function defined by UP 
\begin{equation} \label{sol u2DToda}
  \tau(l, m, n) = \max [ \phi_i (l, m, n+j-1)]_{1\le i, j\le N} ,
\end{equation}
where $\phi_i (l, m, n+j-1)$ is defined by
\begin{equation}  \label{def phi of u2D Toda}
  \phi_i(l, m, n) = \max(\eta_i(l, m, n), \eta '_i(l, m, n))
\end{equation}
with
\begin{equation}  \label{def eta of u2D Toda}
\begin{aligned}
  \eta_i(l, m, n)&=\max (0 , r_i-\delta ) l-\max (0 , \ -r_i-\varepsilon )m +r_i n+c_i, \\
  \eta '_i(l, m, n)&= \max ( 0, \ -r_i-\delta )l- \max ( 0, \ r_i-\varepsilon )m-r_i n+c'_i.
\end{aligned}
\end{equation} 
Here, $r_i$, $c_i$ and $c'_j$ are arbitrary parameters.  In particular, $\phi_i(l, m, n)$ satisfies 
\begin{equation} \label{2dim cond}
\begin{aligned}
  &\phi_i(l+1, m, n) = \max( \phi_i(l, m, n), \ \phi_i(l, m, n+1)-\delta),\\
  &\phi_i(l, m-1, n) = \max( \phi_i(l, m, n), \ \phi_i(l, m, n-1)-\varepsilon ).
\end{aligned}
\end{equation}
Moreover, using the notation $\phi_i(l, m, n+j) \equiv \phi_i(j)$, we have 
\begin{equation} \label{2dim add-cond}
  \phi_{i_1}(j)+\phi_{i_2}(j) \le \max ( \phi_{i_1}(j-1) +\phi_{i_2}(j+1), \ \phi_{i_2}(j-1) +\phi_{i_1}(j+1)) ,
\end{equation}
where $1\le i_1, i_2\le N$.  The above relation gives the reduced expression of tau functions.
\begin{lemma}  \label{lemma4-1}
Tau functions are reduced to 
\begin{align}  
  &\tau(l+1, m, n) =\max_{0\le k_1\le N}( \tau_c(-1, N-k_1)-k_1\delta ),\\
  &\tau(l, m-1, n) =\max_{0\le k_2\le N}( \tau_c(k_2-1, N)-k_2\varepsilon ),  \\
  &\tau(l+1, m, n-1) =\max_{0\le k_1\le N}( \tau_c(N-k_1-1, N)-k_1\delta ), \\
  &\tau(l, m-1, n+1) =\max_{0\le k_2\le N}( \tau_c(-1, k_2)-k_2\varepsilon ), \\
  &\tau(l, m, n) =\tau_c(-1, N), 
\end{align}
and
\begin{equation}
  \tau(l+1, m-1, n) =\max_{0\le k_1, k_2\le N}( \Psi(k_1, k_2)-k_1\delta -k_2\varepsilon ),
\end{equation}
where $\tau_c(\alpha , \beta )$ $(\alpha <\beta )$ is defined by
\begin{equation} 
\tau_c(\alpha, \beta ) = \max [-1 \ \dots \ \widehat{\alpha} \ \dots \ \widehat{\beta} \ \dots \ N].  
\end{equation}
We use $j$ for $(\phi_i(j))_{1\le i\le N}$ and define $\Psi(k_1, k_2)$ as follows:
\begin{equation}  \label{2dToda Psi1}
  \Psi(k_1, k_2) =
 \begin{cases}
  \displaystyle \max_{0\le i\le k_2}(\tau_c(k_2-i-1, N-k_1+i)) & (k_1\ge k_2 \text { and } N-k_1\ge k_2 )\\
  \displaystyle \max_{0\le i\le k_1}(\tau_c(k_2-i-1, N-k_1+i)) & (N-k_1\ge k_2 \ge k_1 )\\
  \displaystyle \max_{0\le i\le N-k_1}(\tau_c(i-1, N-k_1+k_2-i)) & (k_1\ge k_2 \ge N-k_1 )\\
  \displaystyle \max_{0\le i\le N-k_2}(\tau_c(N-k_1-i-1, k_2+i)) & (k_2\ge N-k_1 \text { and } k_2\ge k_1 ).
\end{cases}
\end{equation}
for $1\le k_1, k_2 \le N$.  In the case of $1\le k_1, k_2\le N$, 
\begin{equation} \label{2dToda Psi2}
\begin{aligned}
  \Psi(k_1, k_2) &= \max (\Psi(k_1-1, k_2-1), \ \tau_c(k_2-1, N-k_1)) \qquad &(k_2-1<N-k_1 )\\
  \Psi(k_1-1, k_2-1) &= \max (\Psi(k_1, k_2), \ \tau_c(N-k_1, k_2-1)) \qquad &(k_2-1>N-k_1 )\\
  \Psi(k_1, k_2) &= \Psi(k_1-1, k_2-1) \qquad &(k_2-1=N-k_1 )
\end{aligned}
\end{equation}
hold.  
\end{lemma}
Moreover, we can obtain the following equation by the conditional uPl\"ucker relation.     
\begin{equation}  \label{4-4}
  \tau _c(k_1, N+1)+\tau_c(0, k_2)=\max (\tau _c(k_2, N+1)+\tau_c(0, k_1), \tau _c(0, N+1)+\tau_c(k_1, k_2)),
\end{equation}
where $1\le k_1<k_2<N+1$.  Then, comparing the arguments which have the same $-k_1\delta-k_2\varepsilon$ in 
\begin{equation}  \label{4-2-1}
  \max( \tau(l, m-1, n)+\tau(l+1, m, n), \ \tau(l, m, n)+\tau(l+1, m-1, n)-\delta -\varepsilon ),
\end{equation}
and
\begin{equation} \label{4-2-2}
  \max(\tau(l, m, n)+\tau(l+1, m-1, n), \ \tau(l, m-1, n+1)+\tau(l+1, m, n-1)-\delta -\varepsilon )
\end{equation}
with Lemma \ref{lemma4-1} and (\ref{4-4}), we get Lemma \ref{lemma4-2}.  
\begin{lemma}  \label{lemma4-2}
  The UP (\ref{sol u2DToda}) defined by (\ref{def phi of u2D Toda}) and (\ref{def eta of u2D Toda}) satisfies the equation,
\begin{equation}  \label{u2D Toda2}
\begin{aligned}
  &\max( \tau(l, m-1, n)+\tau(l+1, m, n), \ \tau(l, m, n)+\tau(l+1, m-1, n)-\delta -\varepsilon ) \\
  =&
  \max(\tau(l, m, n)+\tau(l+1, m-1, n), \ \tau(l, m-1, n+1)+\tau(l+1, m, n-1)-\delta -\varepsilon ).  
\end{aligned}
\end{equation}  
\end{lemma}   
Since (\ref{u2D Toda2}) can be reduced to the u2D Toda lattice equation (\ref{u2DToda}), we obtain the following theorem.  
\begin{theorem}
  The UP (\ref{sol u2DToda}) defined by (\ref{def phi of u2D Toda}) and (\ref{def eta of u2D Toda}) satisfies the u2D Toda lattice equation (\ref{u2DToda}).  
\end{theorem}   
\section{Concluding Remarks}
In this article, we consider the specialized UP, and give the conditional uPl\"ucker relation.  Moreover, we show it solves both the uKP and the u2D Toda lattice equation.  Since the determinant solution on continuous or discrete soliton equation are derived from Pl\"ucker relation, the conditional uPl\"ucker relation can be regarded as the ultradiscrete analogue of Pl\"ucker relation.  
However, Pl\"ucker relations used for continuous or discrete soliton equations are quite general formulae on determinants, but strong conditions are necessary for the entry of UP in the uPl\"ucker relation.  In fact, we note there exist a difference  between determinant and UP solutions as below.  The UP solution for the uKP equation (\ref{sol uKP}) is defined by (\ref{def phi of uKP}) and (\ref{def eta of uKP}) and they derive (\ref{uKP cond1}), (\ref{uKP cond1-2}) and (\ref{uKP cond1-3}). 
On the other hand, the discrete KP equation, 
\begin{equation}
\begin{aligned}
  &a_1(a_2-a_3)\tau(l+1, m, n) \tau(l, m+1, n+1)\\
  +& a_2(a_3-a_1)\tau(l, m+1, n) \tau(l+1, m, n+1)\\
  +&a_3(a_1-a_2)\tau(l, m, n+1) \tau(l+1, m+1, n)=0, 
\end{aligned}
\end{equation}
has the determinant solution 
\begin{equation} 
  \tau(l, m, n) = | \varphi_i (l, m, n, s+j-1)|_{1\le i, j\le N} 
\end{equation}
with 
\begin{equation}  \label{disp rel of phi}
\begin{aligned}
  \varphi_i(l+1, m, n, s) =&\varphi_i(l, m, n, s) +a_1\varphi_i(l, m, n, s+1), \\
  \varphi_i(l, m+1, n, s) =&\varphi_i(l, m, n, s) +a_2\varphi_i(l, m, n, s+1), \\
  \varphi_i(l, m, n+1, s) =&\varphi_i(l, m, n, s) +a_3\varphi_i(l, m, n, s+1).
\end{aligned}
\end{equation}
Equation (\ref{disp rel of phi}) corresponds to (\ref{uKP cond1}), (\ref{uKP cond1-2}) and (\ref{uKP cond1-3}).  Then it is expected that the UP solution with only (\ref{uKP cond1}), (\ref{uKP cond1-2}) and (\ref{uKP cond1-3}) also satisfies the uKP equation.  However, it does not.  In fact, for $N=2$, when we set the function $\phi_i(l, m, n, s)$ as
\begin{equation}
\begin{aligned}
  & \phi_1(l, m, n, s) = 10 &  & \phi_2(l, m, n, s)=30\\
  & \phi_1(l, m, n, s+1) = 50 & & \phi_2(l, m, n, s+1) = 0\\
  & \phi_1(l, m, n, s+2) = 0 & & \phi_2(l, m, n, s+2)=40\\
  & \phi_1(l, m, n, s+3) = 100 & & \phi_2(l, m, n, s+3)=0
\end{aligned}
\end{equation}
and $(a_1, a_2, a_3) = (30, 2, 1)$, they satisfy (\ref{uKP cond1}), (\ref{uKP cond1-2}), (\ref{uKP cond1-3}) and also (\ref{uKP cond2}).  Nevertheless, the UP solution provided with the above functions does not satisfy the uKP equation.  Thus, it means the form $|y_i+jr_i|$ is necessary for the UP solution.  It is one of the future problems to clarify the difference between these structures.  
\appendix
\section{Identity of UP's}
We prove an identity of UP's (\ref{id of UP}).  In this appendix, we use the simple notations of the $N\times N$ matrices 
\begin{equation}
\begin{aligned}
  A_j &=  [\bm{a}_1 \ldots \bm{a}_{N-1} \ \bm{b}_j] \quad (1\le j\le 3), \\
  A_{jj'} &=  [\bm{a}_1 \ldots \bm{a}_{N-2} \ \bm{b}_j \ \bm{b}_{j'} ]  \quad (1\le j<j'\le 3),
\end{aligned} 
\end{equation}
and the $(N-1)\times (N-1)$ matrix obtained by eliminating the $k_1$-th row and the $l_1$-th column from $A_j$ as ${A_j} _{k\atop l}$.  In the same way, the $(N-n)\times (N-n)$ matrix obtained by eliminating the $k_1, k_2, \dots ,$ and $k_n$-th rows and the $l_1, l_2, \dots$, and $l_n$-th columns from $A_j$ is denoted by ${A_j}_{k_1, k_2, \dots , k_n \atop{l_1, l_2, \dots , l_n}}$.  These notations give
\begin{equation} \label{A-1-rel1}
  {A_1}_{k_1, \ldots , k_{n-1},  k_n\atop{l_1, \ldots ,l_{n-1},  N}}  ={A_2}_{k_1, \ldots , k_{n-1}, k_n \atop{l_1, \ldots , l_{n-1},  N}} ={A_3}_{k_1, \ldots , k_{n-1}, k_n \atop{l_1, \ldots , l_{n-1}, N}}
\end{equation}
for $1\le l_1< l_2< \dots < l_{n-1}\le N-1$, and 
\begin{equation}  \label{A-1-rel2}
  {A_{23}}_{ k_1, \dots , k_{n-1}, k_n \atop{l_1, \dots , l_{n-1}, N-1}}  ={A_{13}}_{k_1, \dots , k_{n-1}, k_n \atop{l_1, \dots , l_{n-1}, N-1}},
\end{equation}
\begin{equation}  \label{A-1-rel3}
  {A_{23}}_{k_1, \dots , k_{n-1}, k_n \atop{l_1, \dots , l_{n-1}, N}}  ={A_{12}}_{k_1, \dots , k_{n-1}, k_n \atop{l_1, \dots , l_{n-1}, N-1}},
\end{equation}
for $1\le l_1< l_2< \dots < l_{n-1}\le N-2$.  \par
We can expand $\max A_1$  as  
\begin{equation}
  \max A_1 = \max_{1\le k_1\le N} \Bigl( \max {A_1}_{k_1\atop N}+b_{k_11} \Bigr).
\end{equation}
Here $b_{k_11}$ stands for the $k_1$-th element of $\bm{b}_1$.  This expansion corresponds the cofactor expansion.   Similarly, we can derive $\max A_{23}$ by expanding with respect to the $k_1$-th row
\begin{equation}
  \max A_{23} =\max \bigl( \max_{1\le l_1\le N-2}( \max {A_{23}}_{k_1 \atop {l_1}}+ a_{k_1l_1}) , \ \max {A_{23}}_{k_1\atop{ N-1}}+ b_{k_12}, \ \max {A_{23}}_{k_1 \atop N}+ b_{k_13} \bigr).
\end{equation}
for $1\le k_1\le N$.  The symbols $a_{k_1l_1}$, $b_{k_12}$, $b_{k_13}$ mean the $k_1$-th element of $\bm{a}_{l_1}$, $\bm{b}_2$, $\bm{b}_3$ respectively.  Thus, we have
\begin{equation}  \label{A-4}
\begin{aligned}
  \max A_1 +\max A_{23}= \max_{1\le k_1\le N} \bigl( &\max {A_1}_{k_1\atop N}+b_{k_11} +\max_{1\le l_1\le N-2}( \max {A_{23}}_{k_1\atop{ l_1}}+ a_{k_1l_1}),  \\
  &\max {A_1}_{k_1 \atop N}+b_{k_11} +\max {A_{23}}_{k_1 \atop{ N-1}}+ b_{k_12}, \\  
  & \max {A_1}_{k_1 \atop N}+b_{k_11} +\max {A_{23}}_{k_1 \atop N}+ b_{k_13}\bigr).
\end{aligned}
\end{equation}
On the other hand, 
\begin{equation}  \label{A-5}
\begin{aligned}
  \max A_2 +\max A_{13}= \max_{1\le k_1\le N} \bigl( &\max {A_2}_{k_1 \atop N}+b_{k_12} +\max_{1\le l_1\le N-2}( \max {A_{13}}_{k_1 \atop{ l_1}}+ a_{k_1l_1}),  \\
  &\max {A_2}_{k_1 \atop N}+b_{k_12} +\max {A_{13}}_{k_1 \atop {N-1}}+ b_{k_11}, \\  
  & \max {A_2}_{k_1 \atop N}+b_{k_12} +\max {A_{13}}_{ k_1 \atop N}+ b_{k_13}\bigr).
\end{aligned}
\end{equation}
Then using (\ref{A-1-rel1}), the second argument of (\ref{A-4}) is rewritten as 
\begin{equation}
  \max {A_2}_{k_1 \atop N}+b_{k_11} +\max {A_{13}}_{k_1 \atop{ N-1}}+ b_{k_12}.
\end{equation} 
Hence, the second argument of (\ref{A-4}) is equal to that of (\ref{A-5}), in other words, 
\begin{equation}
  \max {A_1}_{k_1 \atop N}+b_{k_11} +\max {A_{23}}_{k_1 \atop{ N-1}}+ b_{k_12} \le \max A_2+\max A_{13}.
\end{equation}
Similarly, it follows that the third argument of (\ref{A-4}) is smaller than or equal to $\max A_3+\max A_{12}$.  Next, let us consider the first argument of (\ref{A-4}), 
\begin{equation} \label{A-6}
  \max {A_1}_{k_1 \atop N} +\max_{1\le l_1\le N-2}( \max {A_{23}}_{k_1 \atop{ l_1}}+ a_{k_1l_1})+b_{k_11}.
\end{equation}
We can derive the first term by expanding with respect to the $l_1(\not= N)$-th column  
\begin{equation} \label{uPlucker6}
  \max {A_1}_{k_1 \atop N} = \max_{1\le k_2\le N, \atop {k_2\not=k_1}}\bigl( \max {A_1}_{k_2, k_1 \atop{ l_1, N}}+a_{k_2l_1}\bigr),
\end{equation}
and the second term with respect to the $k_2$-th row
\begin{equation} \label{A-7}
\begin{aligned}
  &\max_{1\le l_1\le N-2}\bigl( \max {A_{23}}_{k_1 \atop{ l_1}}+ a_{k_1l_1}\bigr)  =\max_{1\le l_1\le N-2}\Bigl( \bigl( \max_{1\le l_2\le N-2 \atop l_2\not=l_1}( \max {A_{23}}_{k_2, k_1 \atop{ l_2, l_1}} +a_{k_2l_2}), \\
  & \max {A_{23}}_{k_2, k_1 \atop{ N-1 , l_1}}+b_{k_22}, \ \max {A_{23}}_{k_2, k_1 \atop{ N, l_1}} +b_{k_23}\bigr)  + a_{k_1l_1}\Bigr).  
\end{aligned}
\end{equation}
Recursively, any argument of $\max A_1+\max A_{23}$ is expressed by either
\begin{equation}  \label{A-8-1}
  \max {A_1}_{k_n, \ldots , k_2, k_1\atop{l_{n-1}, \ldots , l_1,  N}} +\sum_{1\le i\le n-1}a_{k_{i+1}l_i}+b_{k_11}+\max {A_{23}}_{k_n,  k_{n-1}, \ldots , k_1 \atop{N-1, l_{n-1}, \ldots , l_1}} +\sum_{1\le i\le n-1}a_{k_il_i}+b_{k_n2}
\end{equation}
or
\begin{equation}  \label{A-8-2}
  \max {A_1}_{k_n, \ldots , k_2, k_1 \atop{l_{n-1}, \ldots , l_1, N}} +\sum_{1\le i\le n-1}a_{k_{i+1}l_i}+b_{k_11} +\max {A_{23}}_{k_n, k_{n-1}, \ldots , k_1 \atop{N, l_{n-1}, \ldots , l_1}} +\sum_{1\le i\le n-1}a_{k_il_i}+b_{k_n3}.
\end{equation}
Using (\ref{A-1-rel2}) and (\ref{A-1-rel3}), (\ref{A-8-1}) is expressed by
\begin{equation}
  \max {A_2}_{k_{n-1}, \ldots , k_1, k_n \atop{l_{n-1}, \ldots , l_1,  N}}+\sum_{1\le i\le n-1}a_{k_il_i}+b_{k_n2}+\max {A_{13}}_{k_1, k_n, \ldots , k_2, \atop{N-1, l_{n-1}, \ldots , l_1}} +\sum_{1\le i\le n-1}a_{k_{i+1}l_i}+b_{k_11},
\end{equation}
and it is small than or equal to $\max A_2+\max A_{13}$.  We can prove (\ref{A-8-2}) is smaller than or equal to $\max A_3+\max A_{12}$ similarly.  Therefore, we obtain 
\begin{equation} \label{A-1}
  \max A_1 +\max A_{23} \le \max \bigl(  \max A_2 +\max A_{13}, \max A_3 +\max A_{12} \bigr)
\end{equation}
 since any argument of $\max A_1+\max A_{23}$ is smaller than or equal to either $\max A_2+\max A_{13}$ or $\max A_3+\max A_{12}$.  Moreover, 
\begin{equation} 
\begin{aligned}
  &\max A_2 +\max A_{13} \le \max \bigl(  \max A_1 +\max A_{23}, \max A_3 +\max A_{12} \bigr),\\
  &\max A_3 +\max A_{12} \le \max \bigl(  \max A_1 +\max A_{23}, \max A_2 +\max A_{13} \bigr)
\end{aligned}
\end{equation}
also hold from the symmetry, and we get (\ref{id of UP}).   
\section{Proofs of inequalities (\ref{ineq1}) and (\ref{ineq2})}
We prove only (\ref{ineq1}) in this appendix since (\ref{ineq2}) is proved by the similar way.  We note that the idea of the proof is given in \cite{Nakata}.  Let us define $H^N_1$ by
\begin{equation} 
\begin{aligned}
   H^N_1\equiv &\max[ 1 \ \dots \ \widehat{k_2} \ \dots \ N ] + \max[ 2 \ \dots \ \widehat{k_1} \ \dots \ \widehat{k_3} \dots \ N+2 ]\\
  &-\max[ 2 \ \dots \ \widehat{k_2} \ \dots \ N+1 ]- \max[ 1 \ \dots \ \widehat{k_1} \ \dots \ \widehat{k_3} \dots \ N+1 ],
\end{aligned}
\end{equation} 
where $1<k_1<k_2<k_3<N+1$ and $N$ is a natural number satisfying $N\ge 4$.  We use a mathematical induction to prove $H^N_1\le r_N$.  For $N=4$, we can calculate 
\begin{equation} 
   \max[ 1 \ 2 \ 4 ] + \max[ 3 \ 5 \ 6 ] -\max[ 2 \ 4 \ 5 ]- \max[ 1 \ 3 \ 5 ] \le r_4. 
\end{equation} 
Let us suppose $H^N_1\le r_N$ and prove $H^{N+1}_1\le r_{N+1}$.  Using Lemma \ref{lemma3}, we have 
\begin{equation} 
\begin{aligned}
  \max[ 1 \ \dots \ \widehat{k_2} \ \dots \ N+1 ]
  =\max \bigl( &y_N+(N+1)r_N+\max[ 1 \ \dots \ \widehat{k_2} \ \dots \ N ], \\
   &-y_N-r_N+\max[ 2 \ \dots \ \widehat{k_2} \ \dots \ N+1 ] \bigr),
\end{aligned}
\end{equation} 
\begin{equation} 
\begin{aligned}
  \max[ 2 \ \dots \ \widehat{k_1} \ \dots \ \widehat{k_3} \dots \ N+3 ]  =\max  \bigl( &y_N+(N+3)r_N+\max[ 2 \ \dots \ \widehat{k_1} \ \dots \ \widehat{k_3} \dots \ N+2 ], \\
  &-y_N-2r_N+\max[ 3 \ \dots \ \widehat{k_1} \ \dots \ \widehat{k_3} \dots \ N+3 ]\bigr),
\end{aligned}
\end{equation} 
\begin{equation} 
\begin{aligned}
   \max[ 2 \ \dots \ \widehat{k_2} \ \dots \ N+2 ]  =\max \bigl( &y_N+ (N+2)r_N+\max[ 2 \ \dots \ \widehat{k_2} \ \dots \ N+1 ], \\
  &-y_N-2r_N+\max[ 3 \ \dots \ \widehat{k_2} \ \dots \ N+2 ]\bigr),
\end{aligned}
\end{equation} 
\begin{equation} 
\begin{aligned}
   \max[ 1 \ \dots \ \widehat{k_1} \ \dots \ \widehat{k_3} \dots \ N+2 ]  =\max \bigl( &y_N+(N+2)r_N+\max[ 1 \ \dots \ \widehat{k_1} \ \dots \ \widehat{k_3} \dots \ N+1 ],  \\
  & -y_N-r_N+ \max[ 2 \ \dots \ \widehat{k_1} \ \dots \ \widehat{k_3} \dots \ N+2 ] \bigr).
\end{aligned}
\end{equation} 
In the case of $k_1=2$, we define
\begin{equation}
  \max[ 3 \ \widehat{2} \ \dots \ \widehat{k_3} \dots \ N+3 ]
 \equiv -r_N+\max[ 4 \ \dots \ \widehat{k_3} \dots \ N+3 ].
\end{equation}
Here, we have inequalities 
\begin{equation} 
\begin{aligned}
   &\max[ 1 \ \dots \ \widehat{k_2} \ \dots \ N+1 ]-\max[ 1 \ \dots \ \widehat{k_1} \ \dots \ \widehat{k_3} \dots \ N+2 ]\\
  \le \max \bigl( &-r_N+\max[ 1 \ \dots \ \widehat{k_2} \ \dots \ N ]-\max[ 1 \ \dots \ \widehat{k_1} \ \dots \ \widehat{k_3} \dots \ N+1 ], \\
   &\max[ 2 \ \dots \ \widehat{k_2} \ \dots \ N+1 ]-\max[ 2 \ \dots \ \widehat{k_1} \ \dots \ \widehat{k_3} \dots \ N+2 ] \bigr)
\end{aligned}
\end{equation} 
and
\begin{equation} 
\begin{aligned}
   &\max[ 2 \ \dots \ \widehat{k_1} \ \dots \ \widehat{k_3} \dots \ N+3 ]-\max[ 2 \ \dots \ \widehat{k_2} \ \dots \ N+2 ]\\
  \le \max \bigl( &r_N+\max[ 2 \ \dots \ \widehat{k_1} \ \dots \ \widehat{k_3} \dots \ N+2 ]-\max[ 2 \ \dots \ \widehat{k_2} \ \dots \ N+1 ],\\ 
   &\max[ 3 \ \dots \ \widehat{k_1} \ \dots \ \widehat{k_3} \dots \ N+3 ]-\max[ 3 \ \dots \ \widehat{k_2} \ \dots \ N+2 ]\bigr)
\end{aligned}
\end{equation} 
from a formula $\max (x, y)-\max (z, w)\le \max (x-z, y-w)$ for any real numbers $x$, $y$, $z$ and $w$.  Then, a sum of the above inequalities gives
\begin{equation} 
  H^{N+1}_1 \le \max (H^N_1, \ r_N, \ -r_N+H^N_1+H^N_2, H^N_2)\le r_N  
\end{equation} 
for the assumption.  Therefore, we obtain $H^{N+1}_1\le r_{N+1}$.  
\section{Proofs of Lemma \ref{lemma3-3}}
In this appendix, we prove Lemma \ref{lemma3-3}.  The relation (\ref{uKP cond1-2}) derives 
\begin{equation}
\begin{aligned}
  \tau(l, m+1, n+1)&=\max [\phi_i (l, m+1, n+1, s+j-1)]_{1\le i, j\le N}\\
  &=\max_{0\le k_2\le N}( \tau_c(N-k_2, N+1 | n+1)-k_2a_2 ),
\end{aligned}
\end{equation}
where $\tau_c(N-k_2, N+1 | n+1)$ is the same as $\tau_c(N-k_2, N+1)$ except that the label $n$ in $\tau_c(N-k_2, N+1)$ replaced by $n+1$.  Furthermore, applying (\ref{uKP cond1-3}) to each column in $\tau_c(N-k_2, N+1 | n+1)$, we have
\begin{equation}  \label{C-1}
\begin{aligned}
  &\tau_c(N-k_2, N+1 | n+1) = \max [  \max (\bm{\phi}(j-1), \bm{\phi}(j)-a_3) ]_{{\scriptsize 1\le j\le N+1}\atop {\scriptsize j\not= N-k_2+1} } .
\end{aligned}
\end{equation}
Let us consider the maximum of the UP's which have $-k_3a_3$ in (\ref{C-1}).  In the case of $k_3\ge k_2, N-k_2$, for example, it is expressed by
\begin{equation}
\begin{aligned}
  \max\bigl( &\max [0 \ 1 \ \dots \ N-k_3-1 \ \underbrace{N-k_3+1 \ \dots \ N-k_2}_{k_3-k_2} \ \underbrace{N-k_2+2 \  \dots \ N \ N+1}_{k_2}], \\
&\max [0 \ 1 \ \dots \ N-k_3-2 \ \underbrace{N-k_3 \ \dots \ N-k_2}_{k_3-k_2+1} \ N-k_2+1 \ \underbrace{N-k_2+3 \ \dots \ N \ N+1}_{k_2-1}], \\
&\dots ,\\
&\max [\underbrace{1 \ 2 \ \dots \ N-k_2}_{N-k_2} \ N-k_2+1 \ \dots \ 2N-k_2-k_3 \ \underbrace{2N-k_2-k_3+2 \ \dots \ N \ N+1}_{k_3-(N-k_2)}] \bigr)
\end{aligned}
\end{equation}
due to (\ref{property2}) and (\ref{3-3}).  Then, the above is expressed by 
\begin{equation}
 \max_{0\le i\le N-k_3}(\tau_c(N-k_3-i, N-k_2+1+i))  
\end{equation}
and it is equal to $\Psi(k_2, k_3)$ in the case of $k_3\ge k_2, N-k_2$.  We can derive $(\ref{KP Psi1})$ in the other conditions by similar procedure.  \par
The relations (\ref{KP Psi1}) derive (\ref{KP Psi2}).  For example, in the case of $k_2<k_3$, 
\begin{equation}
\begin{aligned}
  \Psi(k_2-1, k_3) =& 
 \begin{cases}
  \displaystyle \max_{0\le i\le N-k_3}(\tau_c(N-k_3-i, N-k_2+2+i)) & (k_3\ge N-k_2+1)\\
  \displaystyle  \max_{0\le i\le k_2-1}(\tau_c(N-k_2+1-k_3+i, N+1-i))
 & (N-k_2+1\ge k_3 )
\end{cases}\\
=&
 \begin{cases}
  \displaystyle \max_{1\le i\le N-k_3+1}(\tau_c(N-k_3-i+1, N-k_2+1+i)) & (k_3\ge N-k_2+1)\\
  \displaystyle  \max_{0\le i\le k_2-1}(\tau_c(N-k_2+1-k_3+i, N+1-i))
 & (N-k_2+1\ge k_3 )
\end{cases}.
\end{aligned}
\end{equation}
On the other hand, 
\begin{equation}
\begin{aligned}
  \Psi(k_2, k_3-1) =& 
 \begin{cases}
  \displaystyle \max_{0\le i\le N-k_3+1}(\tau_c(N-k_3+1-i, N-k_2+1+i)) & (k_3-1\ge N-k_2)\\
  \displaystyle  \max_{0\le i\le k_2}(\tau_c(N-k_2-k_3+1+i, N+1-i))
 & (N-k_2\ge k_3-1)
\end{cases}.
\end{aligned}
\end{equation}
The other relations also hold for the symmetry.  Therefore, we have completed the proofs.  In addition, (\ref{2dToda Psi1}), (\ref{2dToda Psi2}) are also given by the similar procedure.  

\end{document}